\documentclass[aps,superscriptaddress,superbib,twocolumn,groupedaddress]{revtex4}
\usepackage{epsfig}
\usepackage{amsmath,amsthm}
\usepackage[english]{babel}
\RequirePackage{afterpage}
\usepackage[babel=true]{csquotes}
\usepackage{color}

\RequirePackage{tabularx}

\usepackage{pict2e}
\RequirePackage{bm}
\usepackage[final]{neel}
\usepackage{multirow}
\usepackage{ulem}
\graphicspath{{fig/}}

\def \p{P_\mathrm{in}}
\def \q{P_\mathrm{out}}

\def\figurename{Fig.}

\begin{document}
\title{Quantitative analysis of shadow X-ray Magnetic Circular Dichroism Photo-Emission Electron Microscopy}%

\author{S. Jamet}
\affiliation{Univ. Grenoble Alpes, Inst N\'EEL -- F-38000 Grenoble -- France}%
\affiliation{CNRS, Inst N\'EEL, F-38000 Grenoble -- France}%

\author{S. Da Col}
\affiliation{Univ. Grenoble Alpes, Inst N\'EEL -- F-38000 Grenoble -- France}%
\affiliation{CNRS, Inst N\'EEL, F-38000 Grenoble -- France}%

\author{N. Rougemaille}%
\affiliation{Univ. Grenoble Alpes, Inst N\'EEL -- F-38000 Grenoble -- France}%
\affiliation{CNRS, Inst N\'EEL, F-38000 Grenoble -- France}%

\author{A. Wartelle}%
\affiliation{Univ. Grenoble Alpes, Inst N\'EEL -- F-38000 Grenoble -- France}%
\affiliation{CNRS, Inst N\'EEL, F-38000 Grenoble -- France}%

\author{A. Locatelli}
\affiliation{Elettra - Sincrotrone Trieste S.C.p.A., I-34012 Basovizza, Trieste, Italy}%

\author{T. O. Mente\c{s}}
\affiliation{Elettra - Sincrotrone Trieste S.C.p.A., I-34012 Basovizza, Trieste, Italy}%

\author{B. Santos Burgos}
\affiliation{Elettra - Sincrotrone Trieste S.C.p.A., I-34012 Basovizza, Trieste, Italy}%

\author{R. Afid}
\affiliation{Univ. Grenoble Alpes, Inst N\'EEL -- F-38000 Grenoble -- France}%
\affiliation{CNRS, Inst N\'EEL, F-38000 Grenoble -- France}%

\author{L. Cagnon}
\affiliation{Univ. Grenoble Alpes, Inst N\'EEL -- F-38000 Grenoble -- France}%
\affiliation{CNRS, Inst N\'EEL, F-38000 Grenoble -- France}%

\author{S. Bochmann}
\affiliation{Department of Chemistry, Univ. Erlangen, Erlangen, Germany}%

\author{J. Bachmann}
\affiliation{Department of Chemistry, Univ. Erlangen, Erlangen, Germany}%

\author{O. Fruchart}
\affiliation{Univ. Grenoble Alpes, Inst N\'EEL -- F-38000 Grenoble -- France}%
\affiliation{CNRS, Inst N\'EEL, F-38000 Grenoble -- France}%

\author{J. C. Toussaint}
\affiliation{Univ. Grenoble Alpes, Inst N\'EEL -- F-38000 Grenoble -- France}%
\affiliation{CNRS, Inst N\'EEL, F-38000 Grenoble -- France}%

\email[]{olivier.fruchart@neel.cnrs.fr}

\date{\today}

\begin{abstract}
Shadow X-ray Magnetic Circular Dichroism Photo-Emission Electron Microscopy (XMCD-PEEM) is a recent technique, in which the photon intensity in the shadow of an object lying on a surface, may be used to gather information about the three-dimensional magnetization texture inside the object. Our purpose here is to lay the basis of a quantitative analysis of this technique. We first discuss the principle and implementation of a method to simulate the contrast expected from an arbitrary micromagnetic state. Text book examples and successful comparison with experiments are then given. Instrumental settings are finally discussed, having an impact on the contrast and spatial resolution : photon energy, microscope extraction voltage and plane of focus, microscope background level, electric-field related distortion of three-dimensional objects, Fresnel diffraction or photon scattering.
\end{abstract}

\maketitle

Progress is continuous in the decreasing size and increasing complexity of nanosized magnetic systems being designed for either fundamental science or devices. Magnetic microscopies are crucial tools to monitor and understand the properties of such systems. Various types of information are desirable to gather, leading to multiple criteria to classify microscopies: spatial and time resolution, compatibility with environmental parameters such as variable temperature and applied magnetic field, requirements on the sample preparation and compatibility for ex-situ processing such as lithography, correlation with structural information, elemental sensitivity, quantity measured (magnetization, induction, stray field etc.), sensitivity. The most common magnetic microscopies offering spatial resolution below $\unit[50]{\nano\meter}$ and direct sensitivity to magnetization are X-ray Magnetic Circular Dichroism Photo-Emission Electron Microscopy~(XMCD-PEEM)\cite{bib-SCH2002} and (Scanning) Transmission X-ray Microscopy~$[$(S)TXM$]$\cite{bib-FIS2001}, electron holography or Lorentz microscopy\cite{bib-KAS2011,bib-ZWE2007,bib-CHA1999}, Scanning Electron Microscopy with Polarization Analysis~(SEMPA)\cite{bib-ALL2000}, Spin-Polarized Low-Energy Electron Microscopy~(SPLEEM)\cite{bib-BAU1994,bib-BAU2005,bib-ROU2010}.

Yet another criterion is the volume of the sample probed. This criterium is gaining in importance in the context of the emergence of three-dimensional~(3D) magnetic objects and textures. The distribution of magnetization may be truly 3D if along the three directions in space the size of a system lies above magnetic characteristic length scales, such as the dipolar exchange length $\DipolarExchangeLength$ for soft magnetic materials, or the anisotropy exchange length $\AnisotropyExchangeLength$ for a hard magnetic material\cite{bib-HUB1998b}. While this is obviously fulfilled in macroscopic materials, the complexity of magnetic textures is such that it cannot be measured in detail, and besides it cannot be controlled to achieve specific functions. The progress in nanofabrication techniques now allows to design suitable systems, both with top-down and bottom-up approaches. Let us give some examples. Flat magnetic elements are basic building blocks in spintronics, being patterned with great 2D versatility with thin film and lithography technologies. Large lateral dimensions may give rise to 2D magnetic textures, such as the so-called vortex state in disks\cite{bib-SHI2000b}. Such textures are being investigated to design RF oscillator components, relying on the gyrotropic motion of the vortex core driven by a spin-polarized current\cite{bib-ANT2008}. Stacking several disks provides additional degrees of freedom to control the precessional modes or the spectrum purity, for instance allowing to consider vortices with cores aligned parallel or antiparallel\cite{bib-LOC2010}. While coupled nanocubes had been imaged by electron holography in the case of parallel cores\cite{bib-SNO2008}, stacked disks could be imaged recently with magnetic tomography holography, revealing fine details in the case of (repulsive) antiparallel cores\cite{bib-TAN2015}. The topological identity of such structures with so-called merons has been highlighted\cite{bib-PHA2012}. Another example is long and cylindrical nanowires and nanotubes\cite{bib-VAZ2015}, which in a dense array would be the natural geometry to implement a 3D race-track magnetic memory\cite{bib-PAR2008}. Simulation and theory predicted that a new type of domain wall should exist in these systems, with a truly 3D magnetic texture trying to close the flux in all three directions\cite{bib-HER2002a,bib-FOR2002b,bib-THI2006}. Such domain walls are expected to display in their core a Bloch point, an intriguing magnetic object with a local cancellation of magnetization in the otherwise ferromagnetically-ordered material\cite{bib-FEL1965,bib-DOE1968}. Accordingly, they are  named by some\cite{bib-THI2006,bib-KIM2013,bib-FRU2015b} Bloch-point wall. These domain walls have been predicted to be liable to reach high and robust velocities\cite{bib-THI2006}, due to their specific topology providing them with a protection against transformations\cite{bib-FRU2015b}. While domains in tubes\cite{bib-KIM2011b,bib-STR2014} and conventional transverse walls in wires\cite{bib-BIZ2013} had been imaged, recently some of us provided the experimental proof of the existence of Bloch-point walls using shadow XMCD-PEEM\cite{bib-FRU2014} (the technique will be introduced in the next paragraph). Other examples include magnetization processes inside domain walls\cite{bib-FRU2009b,bib-FRU2010a} and dimensional cross-over from vortices to domain walls\cite{bib-FRU2010c,bib-FRU2015}.

\begin{figure}
	\begin{center}\includegraphics{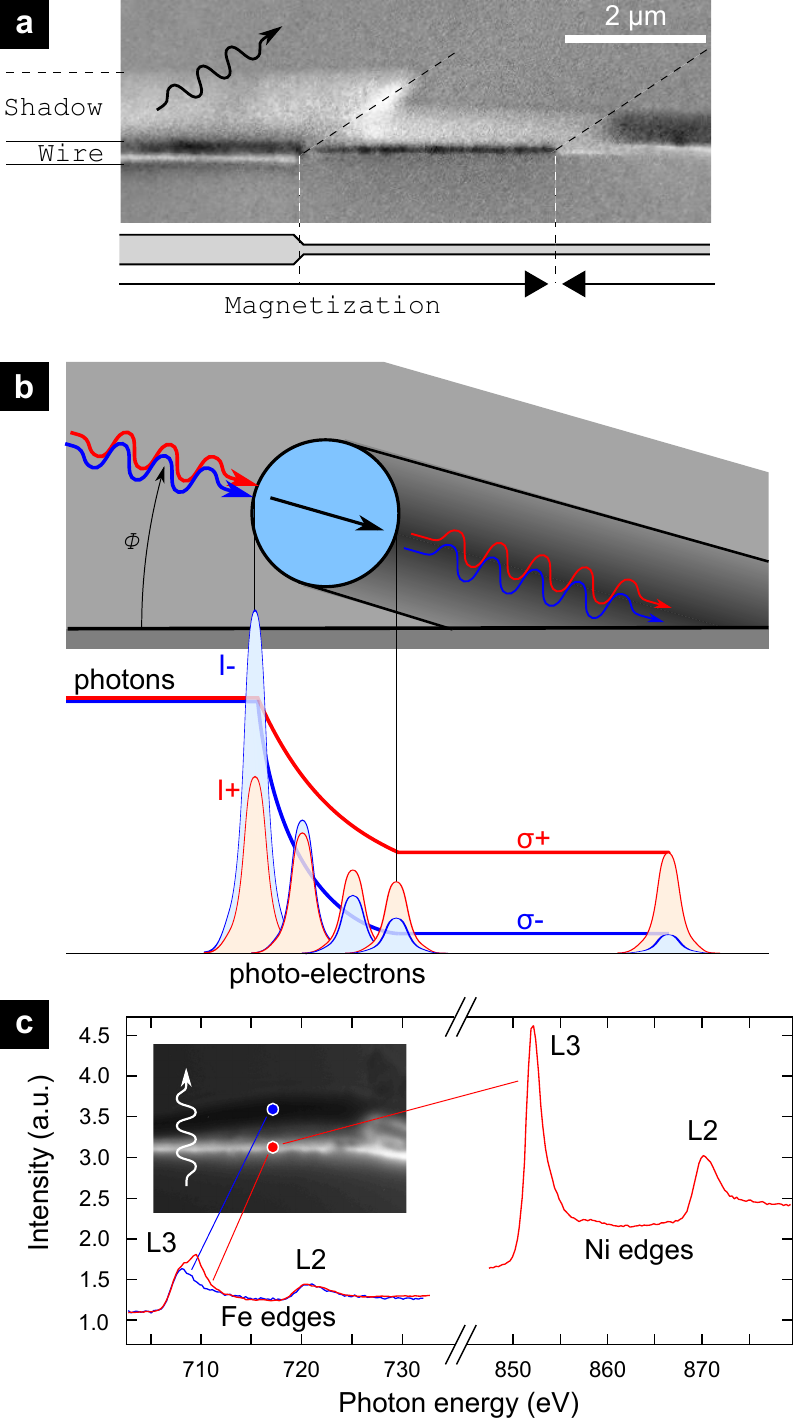}%
 	\caption{\textbf{XAS and XMCD with shadows.} (a)~Shadow XMCD~(top) on a $\mathrm{Fe}_{20}\mathrm{Ni}_{80}$ cylindrical wire with a modulation of diameter, from $\lengthnm{400}$ to $\lengthnm{150}$~(bottom: schematic of wire, the arrows depicting the direction of magnetization). The direction of the beam is indicated by the upper-left wavy arrow. Notice the inversion of surface contrast at the back side of the wire, when its diameter is the largest (left side) (b) Schematic for the dual surface and volume contrast on the basis of the test case of magnetization parallel to the X-ray beam. The curves below represent the polarization-dependent X-ray intensity at the absorption peak as the X-rays propagate through the wire section.  (c) Red: absorption spectra on the wire across the Fe and Ni L edges, normalized to the background signal~(absorption on the supporting Si surface). Blue: inverted and normalized spectra measured in the shadow.\dataref{(a)~2014.03 XMCD-PEEM Elettra 2014.04.02.038}}\label{fig-xas-xmcd}%
	\end{center}
\end{figure}

Let us review again the above-mentioned microscopy techniques, in the light of 3D imaging. SPLEEM and SEMPA typically probe the topmost atomic layer of matter. This makes them sensitive to very small amounts of material if layered, however hides magnetic information in the core of a system. On the reverse, Lorentz, Holography and (S)TXM are transmission techniques with a penetration depth of the order of $\unit[100]{\nano\meter}$, providing information about volume magnetic textures over this depth. However, they have a lower sensitivity as they are not strictly surface-based techniques,  and average the measured signal along the path of the beam. Thus, some information is lost in the case of magnetic textures varying along the depth, unless time- and effort-demanding tomography procedures are implemented\cite{bib-TAN2015}. The probing depth of XMCD-PEEM and SEMPA is intermediate, being a few nanometers and related to the mean free path of the secondary electrons used for imaging. Thus it is not strictly surface sensitive, however not suitable a priori to probe magnetic systems in depth. However, as mentioned above XMCD-PEEM was recently applied to three-dimensional objects lying on a supporting surface\cite{bib-KIM2011b,bib-STR2012b,bib-SAN2012,bib-ZAB2013,bib-FRU2014}. As the X-ray beam is tilted with respect to the normal to the supporting surface, this provides magnetic sensitivity both at the surface of the object, and gives rise to a shadow on the supporting surface, whose inspection yields information about magnetization in the core \bracketsubfigref{fig-xas-xmcd}{a,b}. This has been named shadow XMCD-PEEM  \cite{bib-KIM2011b}. This provides a technique with an interesting hybrid sensitivity, within the set of microscopy techniques mentioned above. However, due to the three-dimensional shape of the objects considered, and the depth- and helicity-dependent absorption of X-rays through the structure, the magnetic contrast cannot simply be interpreted as the projection of magnetization along the direction of the beam, as it is the case for the usual surface XMCD-PEEM. For example, \subfigref{fig-xas-xmcd}{a} shows that the contrast at the surface of a uniformly-magnetized 3D object may vary and even change in sign, depending on it size. In Ref.\cite{bib-STR2014} the authors simulated the contrast in the shadow of a rolled tube. However, it was based on an analytical form for the distribution of magnetization in a thin sheet, not a simulated micromagnetic configuration. Also, the contrast at the surface of the structure was not computed.

In this manuscript we review specific aspects of shadow XMCD-PEEM, and propose a method to analyze the resulting images of surface and shadow based on the complete micromagnetic structure of an object, to make shadow XMCD-PEEM a quantitative technique. The manuscript is organized as follows. First the principles and implementation of a method to simulate the contrast of three-dimensional magnetization textures are described. Then we illustrate the simulations with two test cases. Comparison with a few experimental cases is then made, followed by a discussion of the considerations required for the quantitative analysis of magnetic contrast and spatial resolution. These considerations are largely related to finer points of the physics at play, which have so far been left aside in the modeling.


\section{Method and implementation of the simulations}
\label{sec-method}

\begin{figure}
	\begin{center}%
  \includegraphics{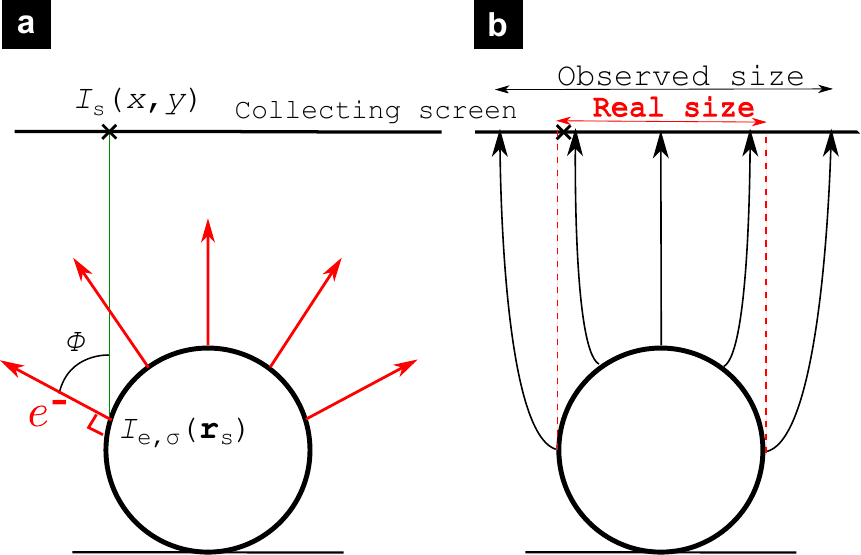}%
  \caption{\textbf{Collecting the photoelectrons}. (a)~Illustration of electrons escaping the magnetic object. $\vect e^-$ is the secondary electron emission direction, and $\theta$ is the angle between the imaging axis and the latter direction. (b)~Illustration of the distortion of the photoelectron trajectory due the electric field extracting electrons into the microscope column (see \protect{\secref{sec-resolution}}).}
\label{fig-scheme-coll}%
  \end{center}
\end{figure}

\begin{table}
  \begin{center}
      \caption{Absorption coefficients $\mu$ of Fe, Ni and $\mathrm{Fe}_{20}\mathrm{Ni}_{80}$ for the photon energy set at the L absorption edges of either Fe and Ni. Figures for pure elements are derived from Ref.\citenum{bib-NAK1999}. Figures for the alloy are linear interpolation of figures for single elements. }
	\label{tab-coeffabs}

\begin{tabularx}{0.85\linewidth}{p{0.16\linewidth}p{0.16\linewidth}p{0.16\linewidth}p{0.16\linewidth}p{0.16\linewidth}p{0.21\linewidth}}
\hline\hline
Edge &$\mu $ ($\reciprocal{\nano\meter}$)& Fe & Ni &Fe$_{20}$Ni$_{80}$\\
\hline
\multirow{2}{*}{Fe L2}&$\mu_{-}$ & 0.03 & $\approx 0$& 0.006\\
&$\mu_{+}$ & 0.04 & $\approx 0$& 0.008\\
\multirow{2}{*}{Fe L3}  &$\mu_{-}$ &0.09 & $\approx 0$& 0.018\\
&$\mu_{+}$ & 0.05 & $\approx 0$& 0.010\\
\multirow{2}{*}{Ni L2}  &$\mu_{-}$ & 0.017& 0.017& 0.017\\
&$\mu_{+}$  & 0.017 & 0.021 &0.020\\
\multirow{2}{*}{Ni L3}  &$\mu_{-}$ &  0.017& 0.053 & 0.046\\
&$\mu_{+}$ &  0.017 &0.040& 0.035\\\hline\hline 		\end{tabularx}
 	\end{center}
\end{table}

Our approach consists in considering a given three-dimensional magnetization texture in a system, and use it as an input to simulate the X-ray absorption spectroscopy~(XAS, arising from a non-polarized photon beam) and XMCD contrasts~(the asymmetry ratio arising from photons with opposite circular polarizations) expected at both the surface and in the shadow. The magnetic texture may be a simple analytical form, or a realistic distribution of magnetization resulting from a micromagnetic simulation. In this section we first describe the physical principles considered to convert a magnetic texture into a magnetic contrast. Then we detail the practical implementation in the numerics.

\subsection{Principle of the method}

Building a XMCD-PEEM image requires to describe mainly three distinct steps including physical and instrumental aspects: absorption through matter; photo-emission of electrons close to surfaces; collection of these electrons in the microscope. The way we model each of these processes is detailed below.

\subsubsection{X-ray absorption}

At any stage when traveling through matter, an X-ray beam is associated with a probability of absorption per unit length, $\mu$, determining the mean free path of photons~$\lambda = 1/\mu$. These parameters depend on the composition of matter, as well as on the photon energy. When comparing the model with experiments we will consider objects made of permalloy, with composition $\mathrm{Fe}_{20}\mathrm{Ni}_{80}$. This will give us the opportunity to discuss the case of alloys, and highlight that this gives more freedom for the choice of the absorbtion edge. We considered photons at the L3 and L2 edges of both Fe and Ni, and used the experimentally-determined parameters from the literature for pure elements and both helicities of photons\cite{bib-NAK1999}. Based on the very similar volume density of these elements and alloys, we assumed that the absorption coefficient of permalloy is $\mu=0.2\mu_\mathrm{Fe}+0.8\mu_\mathrm{Ni}$. Absorption coefficients at the different edges are summarized in \tabref{tab-coeffabs}. At the Fe L edges, absorption due to Fe is large, and the pre-edge absorption of Ni is weak, so that despite the low concentration of Fe it is by far the dominating contribution. At the Ni L edges the post-edge absorption of Fe is in principle no more negligible especially because absorption on Ni with a nearly filled $3d$ band yields a moderate intensity. The contribution of Ni is however still larger to that of Fe, due to its much larger concentration.

The progressive absorption of the beam in matter is described by integrating its position-sensitive rate of absorption through each elementary segment with length $\mathrm{d}\ell$:
\begin{equation}\label{eq-1}
 \frac{\mathrm{d}I_{X,\sigma_{\pm}}}{\mathrm{d}\ell}=-\left[ \frac{1}{2}\mu_+(1\pm   \mathbf {\hat k} \cdot \vect m) + \frac{1}{2}\mu_-(1\mp \mathbf {\hat k} \cdot \vect m)\right]I_{X,\sigma_{\pm}}
\end{equation}%
\noindent where $\mu_+$ and $\mu_-$ stand for the absorption coefficients for left and right circularly polarized X-rays, respectively. This formula takes into account the energy and helicity dependence, in relation with the direction of magnetization in the sample, with $ \mathbf{\hat k}$ the unit vector along the propagation direction.

\subsubsection{Local emission of electrons}

We need to estimate the local rate of emission of photoelectrons $I_{\mathrm{e},\sigma_\pm}(\mathbf{r}_\mathrm{s})$ at any location $\mathbf{r}_\mathrm{s}$ at the surface, resulting from the transmitted X-ray intensity $I_{X,\sigma_{\pm}}(\mathbf{r}_\mathrm{s})$ reaching that location, as calculated previously. The majority of emitted current consists of secondary electrons, whose escape depth is only a few nanometers. As this length is much smaller than the size of objects of interest in shadow-PEEM, and also smaller than any magnetic length scale, we used the simplifying assumption that on the object $I_{\mathrm{e},\sigma}(\vect r_\mathrm{s})$ reflects $I_{X,\sigma}(\vect r_\mathrm{s})$ and magnetization at the surface, through again the dichroism ratio depending on the photon helicity. To the contrary, when impinging on the non-magnetic surface, for instance in the shadow, the photons give rise to a rate of electrons directly proportional to $I_{X,\sigma}(\vect r_\mathrm{s})$. Let us finally discuss the escape of electrons from matter. Initially, photo-emitted electrons are emitted isotropically and not perpendicular to the local sample surface, implying some lateral broadening of the electron emission. Thus, we expect that a measured image results from the convolution of the signal described aforehand, with a function describing these processes. In practice however, as the escape depth of electrons is only a few nanometers, the expected broadening should not exceed these few nanometers, which is much smaller than the instrumental resolution (circa \lengthnm{30}). Thus, these effects may be safely neglected. So, at this stage we have an estimate $I_{\mathrm{e},\sigma}(\vect r_\mathrm{s})$ of the local electron emission at each point of the nanostructure and its supporting surface.

\subsubsection{Intensity on the detector}

We now need to convert the local emission rate $I_{\mathrm{e},\sigma}(\vect r_\mathrm{s})$, into the intensity per unit surface $I_\mathrm{s}(x,y)$, on the detector~(the subscript \textsl{s} standing for \textsl{screen}). One important parameter to make the link between $I_{\mathrm{e},\sigma}(\vect r_\mathrm{s})$ and $I_\mathrm{s}(x,y)$ is the angular acceptance of the microscope. A key parameter is the contrast aperture, whose aim is to select electrons escaping the sample essentially along the column axis, in order to minimize aberrations. While this has no impact for flat surfaces (as for the contrast in the shadow), it affects the contrast arising from 3D objects. Indeed, the emission of secondary electrons is maximum along the normal to the local surface, and the total collecting efficiency is lower for tilted surfaces\cite{bib-MEN2012}. However, the exact angular dependance is not well characterized and may depend sensitively on extraction and electron energy, aperture, surface roughness etc. Thus we did not attempt to consider a realistic transfer function, as this would be largely arbitrary. We simply considered that $I_\mathrm{s}(x,y)=I_{\mathrm{e},\sigma}(\vect r_\mathrm{s})$. To understand which transfer function this stands for, let us call $\theta$ the angle between the microscope axis and the normal to the emitting surface\bracketsubfigref{fig-scheme-coll}{a}. The ratio of local sample surface over corresponding detector surface scales like $1/\cos\theta$, so that if all emitted electrons were collected irrespective of their direction when escaping the surface, a similar ratio would be calculated on the detector. Thus, the choice $I_\mathrm{s}(x,y)=I_{\mathrm{e},\sigma}(\vect r_\mathrm{s})$ is equivalent to using a $\cos\theta$ collection function. Notice that while the collection function affects the XAS, theoretically XMCD images should not depend on this function as they are computed as differences normalized by the sum. In practice, due to the reduced number of photons in areas where the real transfer function does not allow the collection of electrons, combined with a background electron level to be discussed in \secref{sec-bcglevel}, in experimental images the XMCD may be sharply decreased in such areas.

Let us note that the above procedure is not a bijection but a surjection, because of the integration along a path and also of the projection of magnetization along the beam. Thus, one XPEEM image may in principle correspond to different magnetic configurations.  This, along with other issues contributing to image formation such as photon scattering, field distortion due to the object topography, and background electron intensity, will be discussed further below (\secref{sec-contrast}).

\subsection{Numerical implementation}

\begin{figure*}
	\begin{center}%
  \includegraphics{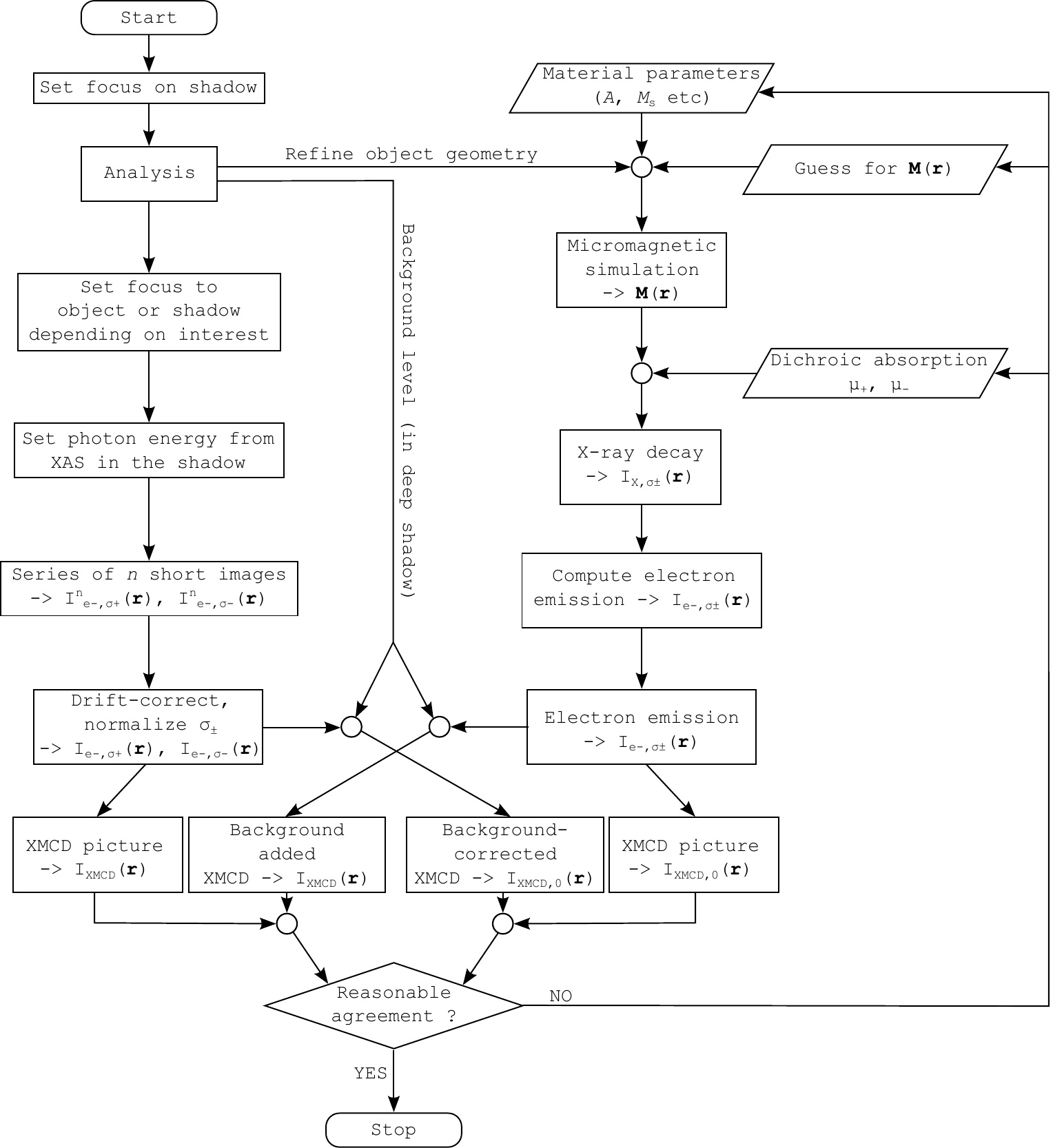}%
  \caption{\textbf{Flow chart of the analysis method}. Key acquisition and simulation steps for the quantitative analysis of shadow XMCD-PEEM images.}
\label{fig-flowchart}%
  \end{center}
\end{figure*}

Besides analytic test cases, micromagnetic configurations resulting from simulations are used as input to compute the XMCD contrast. For the micromagnetic configurations, we use the home-made code FeeLLGood\cite{bib-ALO2012}. FeeLLGood is based on the finite element method. We used material parameters suitable for permalloy: $\mathrm{A}=\unit[10]{\pico\joule\per\meter}$, $\mu_0 M_{\mathrm{s}}=\unit[1]{T}$. The damping parameter $\alpha$ was set to 1 to facilitate convergence, with no impact on the results as we only consider states at rest. No magnetocristalline anisotropy was considered. A flow chart of experimental and simulation steps is provided in \figref{fig-flowchart}.

We illustrate here the simulation scheme with cylindrical nanowires, although it can be applied to an arbitrary shape. The principle of the numerical method is to consider a ray, and to compute the absorption along this ray taking into account the distribution of magnetization in the sample.

\subsubsection{X-ray absorption}

\begin{figure}
	\begin{center}\includegraphics{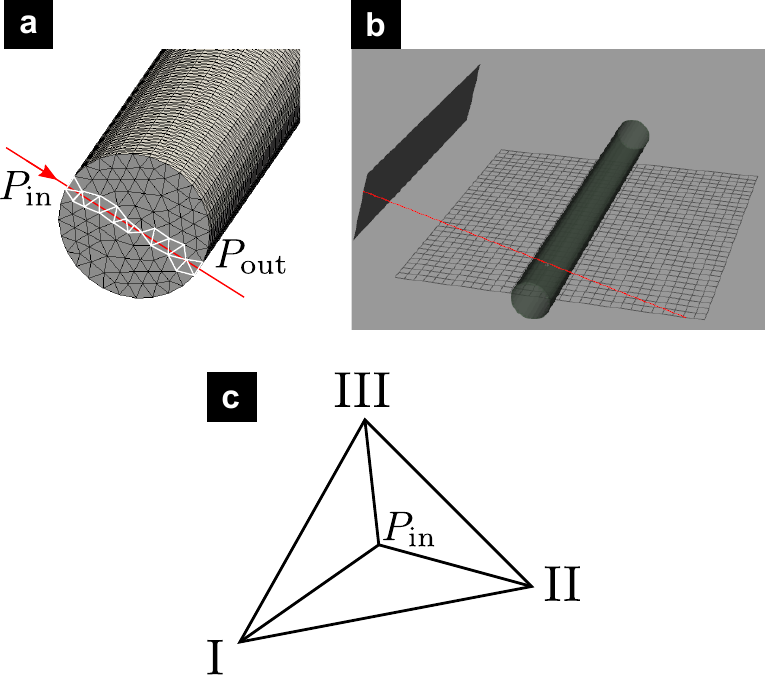}%
 	\caption{\textbf{Description of interpolation methods.} The system considered as an example is a cylindrical nanowire. (a) presents the method to model the X-ray beam. The red line stands for the X-ray beam that crosses the tetrahedron elements (in white, from the finite element discretization). (b) shows the photon beam, and the wire on the regularly-gridded supporting surface. (c) scheme of a triangle to illustrate in 2D the method of the areal coordinates for interpolating magnetization along each ray.}\label{fig-integration}%
	\end{center}
\end{figure}

We first simulate the photon flux $I_{X,\sigma}(\vect r)$ with a given helicity $\sigma$ at any position in space, inside and outside the magnetic system. For this we consider a regular planar grid perpendicular to the direction of the photons of an incident plane wave. From each node a ray is launched and intercepts the surface of the cylinder at two points $\p$ and  $\q$ respectively, called the entering and exit points. In order to calculate the absorption along the ray the latter is discretized into segments of a given length between $\p$ and $\q$. At any of these points we estimate magnetization using an interpolation method~(see \figref{fig-integration}). At $\p$ and $\q$ the magnetization $\vectM$ is interpolated from its values known at the nodes of the triangle (the face of the first/last tetrahedron) to which it belongs, using a method based on the areal coordinates method \cite{bib-COX1969}: $\vect M=\sum_i \alpha_i \vect M_i$, with $i=\mathrm{I},\mathrm{II},\mathrm{III}$ and $\alpha_\mathrm{I}(P_\mathrm{in})= S(P_\mathrm{in}, \mathrm{II}, \mathrm{III})/S(\mathrm{I},\mathrm{II},\mathrm{III})$\bracketsubfigref{fig-integration}c. $S(P_\mathrm{in}, \mathrm{II},\mathrm{III})$ is the surface of a sub-triangle and $S(\mathrm{I},\mathrm{II},\mathrm{III})$ is the total surface of the triangle. $\alpha_\mathrm{II}$ and $\alpha_\mathrm{III}$ are obtained upon circular permutation. For points inside the sample, we first determine the corresponding element (tetrahedron in the volume of the wire, see \subfigref{fig-integration}{a}). The magnetization is then interpolated with the method previously described but with four nodes and making use of sub-tetrahedrons instead of sub-triangles. Then, the progressive absorption along the entire path through the magnetic structure can be computed. It is done by integrating \eqnref{eq-1}: $I_{X,\sigma_{\pm}}\propto \exp\left\{-\int^{\q}_{\p} \mathrm{d}\ell\left[\frac{1}{2}\mu_{+}(1\pm   \mathbf {\hat k} \cdot \vect m) + \frac{1}{2}\mu_{-}(1\mp \mathbf {\hat k} \cdot \vect m)\right]\right\}$.

\subsubsection{Local emission of electrons}

The next step is to get the intensity of the photoelectrons emitted after the absorption of X-rays. At the surface of the wire, the intensity of the photoelectrons is then calculated as the intensity of the photons multiplied by a pre-factor and normalized by the density of X-ray flux $\vect n \cdot \mathbf {\hat k}$. The pre-factor takes into account the scalar product of the local magnetization and the wave vector and the local absorption coefficient such that :
\begin{multline}
I_{\mathrm{e},\sigma_\pm} \propto \frac{1}{\vect n \cdot \mathbf {\hat k}}\;\frac{2}{\mu_+ + \mu_-}\left[\frac{1}{2}\mu_{+}(1\pm   \mathbf {\hat k} \cdot \vect m) \right .  \\
 \left .+ \frac{1}{2}\mu_{-}(1\mp \mathbf {\hat k} \cdot \vect m)\right]I_{X,\sigma_{\pm}.}
\end{multline}%
\noindent Note that the right hand side is the derivative of X-ray intensity along the propagation path [as in \eqnref{eq-1}]. The physical meaning is clear, because the derivative of $I_{X,\sigma}$ is indicative of photons absorbed within that incremental distance, and photoemission is directly proportional to absorption.

\subsubsection{Intensity on the detector}

The last step is to infer the electron intensity on the detector on a square grid, from the photoelectron intensity previously computed at points of the sample and supporting surface. This is achieved with a linear interpolation, in a similar fashion as described above for $\p$ and $\q$. Finally, the XMCD-PEEM contrast is computed as ($I_{\mathrm{e},\sigma_-}-I_{\mathrm{e},\sigma_+})$ and normalized to the sum $(I_{\mathrm{e},\sigma_-}+I_{\mathrm{e},\sigma_+}$). The implementation has been done by using the geometry library CGAL\cite{bib-CGAL} for the use of rays, and the nearest neighbor searching library ANN\cite{bib-ANN}.

\section{Illustration on test cases}
\label{sec-test-cases}

In this section we apply the simulation method to two test cases of analytical distributions of magnetization: transverse uniform magnetization and orthoradial curling\bracketfigref{fig-test-cases}. Curling is a long-existing name is magnetism, used to describe an area where the curl of magnetization is non-zero, such as around a magnetic vortex, or in an object subject to the so-called curling nucleation mode as initially introduced\cite{bib-FRE1957}. Although they would not occur in wires as such, these distributions are chosen to illustrate the method, and understand special features of contrast which can arise in shadow XMCD-PEEM. They are also relevant for the experimentally observed magnetic structures observed in cylindrical wires as we will show in the following sections. These two situations have been described analytically for each point, and we checked that an excellent agreement was found with the numerical grid method. In \figref{fig-test-cases}, the simulated wire was suspended above the substrate surface so that the complete shadow is collected. Although this may happen experimentally in some cases\bracketsubfigref{fig-bpw}{b}, in most cases the wire is in contact with the supporting surface so that part of the shadow is not visible on the screen.

\subsection{Transverse uniform magnetization}
\subfigref{fig-test-cases}{a} shows $I_{X,\sigma\pm}(\vect r)$ and the resulting dichroic absorption for photons going through a wire uniformly magnetized along its diameter. The two curves illustrate the fact that photons with one of the two polarizations is more absorbed than the other, due to magnetic dichroism\bracketsubfigref{fig-xas-xmcd}{b}. The dichroic contrast in the shadow is therefore opposite to that at the front surface  of the wire, as it simply reflects the effect in the transmitted photons. Besides, even though photo-emission is higher~(resp. lower) per photon, for the photons with lower~(resp. higher) transmission, the dichroism measured at the back side of the wire may be reversed compared to the front side, for thick-enough wires and thus large imbalance after transmission. The dichroic contrast at the surface of the wire is illustrated in \subfigref{fig-test-cases}{a}, constant on the front side and gradually decreasing on the back side. The critical diameter above which the contrast reverses may be calculated:

\begin{equation}
d_\mathrm{c} = \frac{\ln(\mu_+/\mu_-)}{\mu_+-\mu_-}\;\frac1{\sqrt{1-\sin^2 \phi}}
\end{equation}%
\noindent where $\phi$ is the incidence angle of the photons\bracketsubfigref{fig-xas-xmcd}a. Obviously, $d_\mathrm{c}$ depends on the X-ray energy via the absorption coefficient $\mu_\pm$. In the case of a wire made of permalloy, and for a grazing angle $\phi=\unit[16]{\degree}$ as will be reported for experimental comparison, $d_\mathrm{c}$  is respectively 70, 140, 20 and \unit[50]{nm} at the Fe-L3, Fe-L2, Ni-L3 and Ni-L2 edges. This explains the dominating inverted contrast on the wide-diameter side of the wire in \figref{fig-xas-xmcd}a.  Note also that the contrast is expected to be larger at the center of the shadow than at its border, because the length of material probed is larger, and so does the imbalance of outgoing photons. These facts highlight that the contrast does not reflect directly the local direction of magnetization, and stresses the need for simulation. Practical examples will be provided in \secref{sec-contrast}, dedicated to the analysis of contrasts.

\subsection{Orthoradial curling}

\begin{figure}
	\begin{center}			\includegraphics{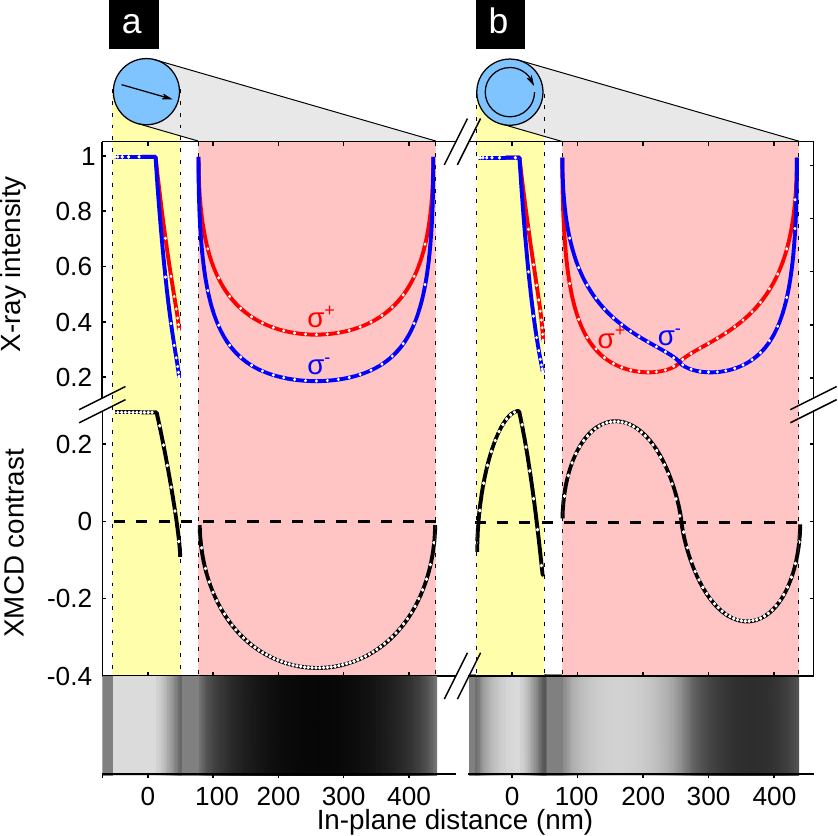} 		\caption{\textbf{Illustration of XMCD-PEEM post-processing on two test cases.} The first test case is (a) uniform magnetization across the wire and parallel to the X-ray beam; the second case (b) is orthoradial curling. The lower part presents the photon density and XMCD for each case at the surface of the wire (yellow background) and in the shadow (pink background). The modeled wire is suspended above the surface, so that the entire shadow is visible. Note that the lateral scale is expanded by a factor $\sin(\angledeg{16})\approx3.6$ in the shadow, thanks to the grazing incidence.}
\label{fig-test-cases}%
  \end{center}
\end{figure}

The case of orthoradial curling of magnetization~(see \subfigref{fig-test-cases}b) is directly relevant for one type of domain wall in cylindrical wires: the Bloch point wall \cite{bib-FRU2014, bib-FRU2015b}. At the bottom part of the wire, magnetization is mostly pointing left, while at the top part it is mostly pointing right. This leads to opposite contrasts on either side of the shadow. The center of the shadow has no XMCD contrast, as at all points through the wire diameter the X-ray direction is perpendicular to the magnetization direction. This dipolar contrast is a clear signature of orthoradial curling. To the contrary, the dichroic contrast at the surface of the wire is maximum close to its top, where the beam is tangent to its surface. It decays on both sides, with a slight negative value on the front side due to the tilted incidence of the photons, and a possible inversion of contrast on the backside depending on the total absorption. Thus, the contrast is largely monopolar as in the case of uniform transverse magnetization, which could for instance naively be expected from a transverse wall with the transverse component aligned with the beam direction. Thus, ascribing the surface contrast to a transverse wall or a Bloch-point wall may remain ambiguous. This example shows that inspection of the shadow may be crucial to get information about a three-dimensional configuration of magnetization.

\section{Comparison with experiments}\label{sec-comparison}

To put the model into practice, we take the example of experiments using shadow XMCD-PEEM on cylindrical wires, consistent with the test cases discussed above. In this section, we describe the experimental setup, and discuss the correspondence of the experimentally observed images with those from the simulations.

\subsection{Experimental details}

The samples considered are permalloy cylindrical wires electroplated in self-organized anodized alumina templates. The alumina matrix is dissolved and wires are dispersed on a naturally-oxidized Si supporting surface. The wires are aligned along a preferential in-plane direction thanks to an in-plane magnetic field applied during dispersion. Their diameter, possibly modulated along the length, ranges from $\unit[50]{\nano\meter}$ to several hundreds of~nm. The length of the wires is typically a few micrometers\cite{bib-FRU2014}.

\noindent Element-selective XAS and XMCD-PEEM were carried out at the spectroscopic photoemission and low-energy electron microscope\cite{bib-LOC2006} operated at the undulator beamline Nanospectroscopy of Elettra, Sincrotrone Trieste. The photons impinge on the surface with a grazing angle $\phi=\unit[16]{\degree}$. Spectroscopy was performed across the L edges of either Ni or Fe, using elliptically-polarized radiation as a probe. Series of several tens of images with an exposure time of few seconds are recorded, drift-corrected and finally co-added. This yields a high signal-over-noise ratio while limiting drift effects, providing images with a spatial resolution of the order of $\unit[30]{\nano\meter}$. The level of circular polarization at the Fe and Ni L~edges was estimated to be around $\unit[75]{\%}$ as the X-ray beam is produced by a higher harmonic of the undulator source.

\subsection{Experimental test cases: curling structures}

\begin{figure}
	\begin{center}%
        \includegraphics{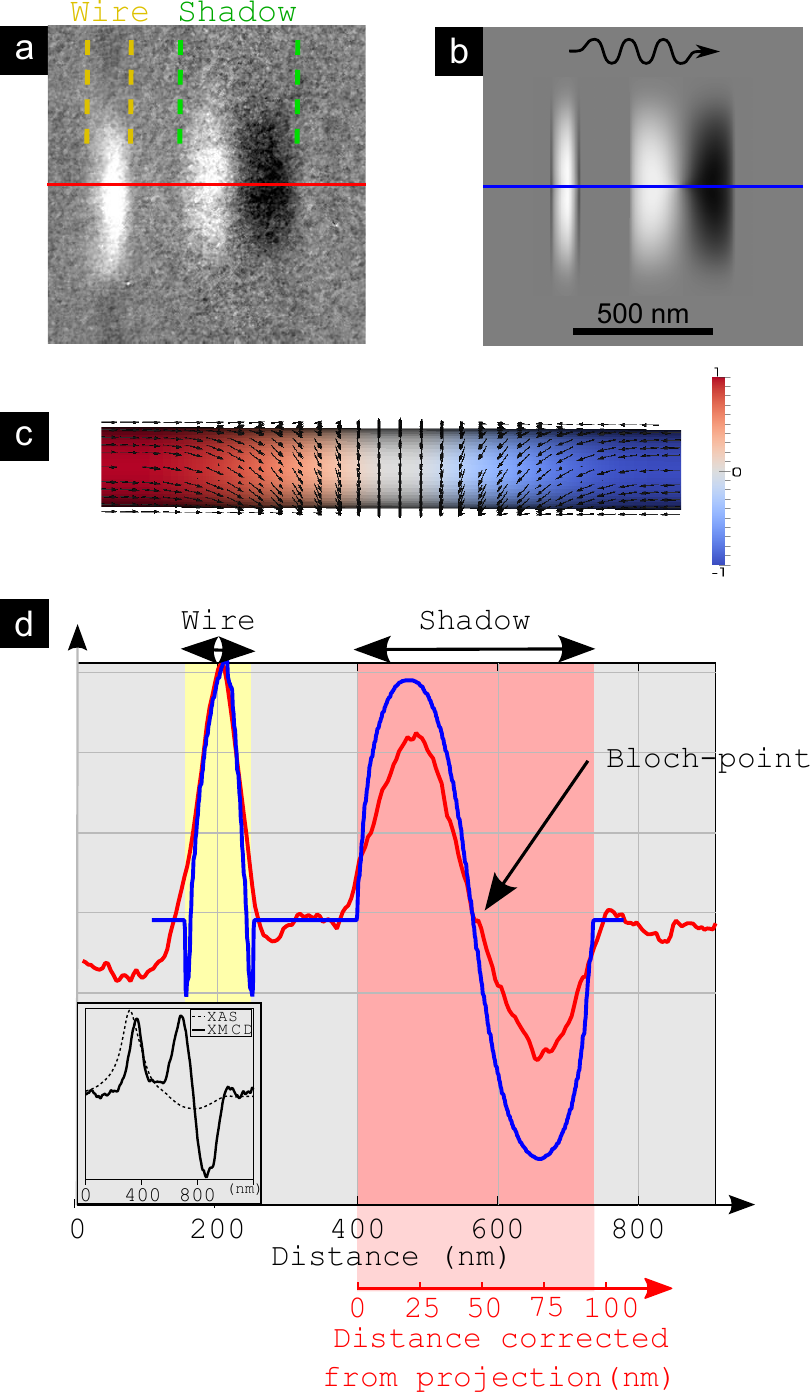}%
  	\caption{\textbf{Comparison of experiment and simulation for the Bloch-Point wall.} (a)~Experimental and (b)~simulated contrasts at the surface and in the shadow (resp. left and right in both images) at the Fe L3 edge for a wire of diameter $\unit[90]{\nano\meter}$. This diameter has been determined from the XAS of the wire shadow. (c)~Micromagnetic simulation of a Bloch-point wall, used as an input for simulating the contrast (d)~Cross-sections for (a) and (b). The insert shows the XAS~(dotted line) and XMCD~(full line) spectra  for the experimental contrasts (a).  }%
 	\label{fig-bpw}%
  \end{center}
\end{figure}

Two types of domain walls may be expected in cylindrical nanowires: of mixed transverse-vortex type for diameter below typically $7\DipolarExchangeLength$ ($\DipolarExchangeLength=\sqrt{2A/\muZero\Ms^2}$), and of Bloch-point type for diameter above typically $7\DipolarExchangeLength$\cite{bib-FRU2015b, bib-THI2006}. The former is reminiscent of domain walls already known in flat strips\cite{bib-MIC1997,bib-NAK2005}, while the latter is specific to wires with large dimensions. Consistently, we observed two types of contrast for domain walls in nanowires, which we could ascribe to these walls\cite{bib-FRU2014}. Here we illustrate the shadow technique with the Bloch-point wall~\bracketfigref{fig-bpw}.

\subfigref{fig-bpw}{a} shows the XMCD-PEEM image from a magnetic wire of diameter \unit[90]{\nano\meter}, suspended above the substrate in the imaged region. A good agreement is found with the contrast delivered by our model\bracketsubfigref{fig-bpw}{b}, based on the simulated micromagnetic configuration\bracketsubfigref{fig-bpw}{c}. Thus, this domain wall can be unambiguously ascribed to a Bloch-point wall. The striking feature is the dual bright and dark contrasts in the shadow, revealing a local orthoradial curling as already seen in \subfigref{fig-test-cases}{b}. The symmetry with respect to a plane perpendicular to the wire axis shows that curling is purely orthoradial, which is compatible only with the Bloch-point wall\cite{bib-FRU2015b}. A quantitative comparison may be made with a cross-section\bracketsubfigref{fig-bpw}{d}. The contrast has been normalized so that the maxima coincide at the surface of the wire. Compared to simulations, the experimental cross-section is wider by approximately $\unit[25]{\nano\meter}$ on either side, which is consistent with the expected instrumental spatial broadening. The agreement is however excellent at the surface of the wire, especially the rather sharp maximum and its location away from the central part and towards the backside of the wire\bracketsubfigref{fig-bpw}{d, insert}. This feature is explained as follows. The XAS should have the shape shown on the top part of \subfigref{fig-test-cases}b. Upon convolution with the resolution function, this initially asymmetric XMCD shape gives rise to a maximum displaced towards the backside of the wire. Note also that in the shadow, the cross-section is clearly antisymmetric, as expected. The cancellation of contrast at the core of the shadow should coincide with the location of the Bloch point. The experimental contrast is however lower than expected in theory, which will be discussed in \secref{sec-contrast}. Of importance is the fact that the lateral scale of the wire is expanded by a factor $1/\sin(\angledeg{16})\approx3.6$ in the shadow, thanks to the grazing incidence. In principle this promises an increase of spatial resolution of 3.6 along one direction, however issues of signal-over-noise ratio may limit this gain, which will be addressed in the next section.

\begin{figure}
	\begin{center}%
  	\includegraphics{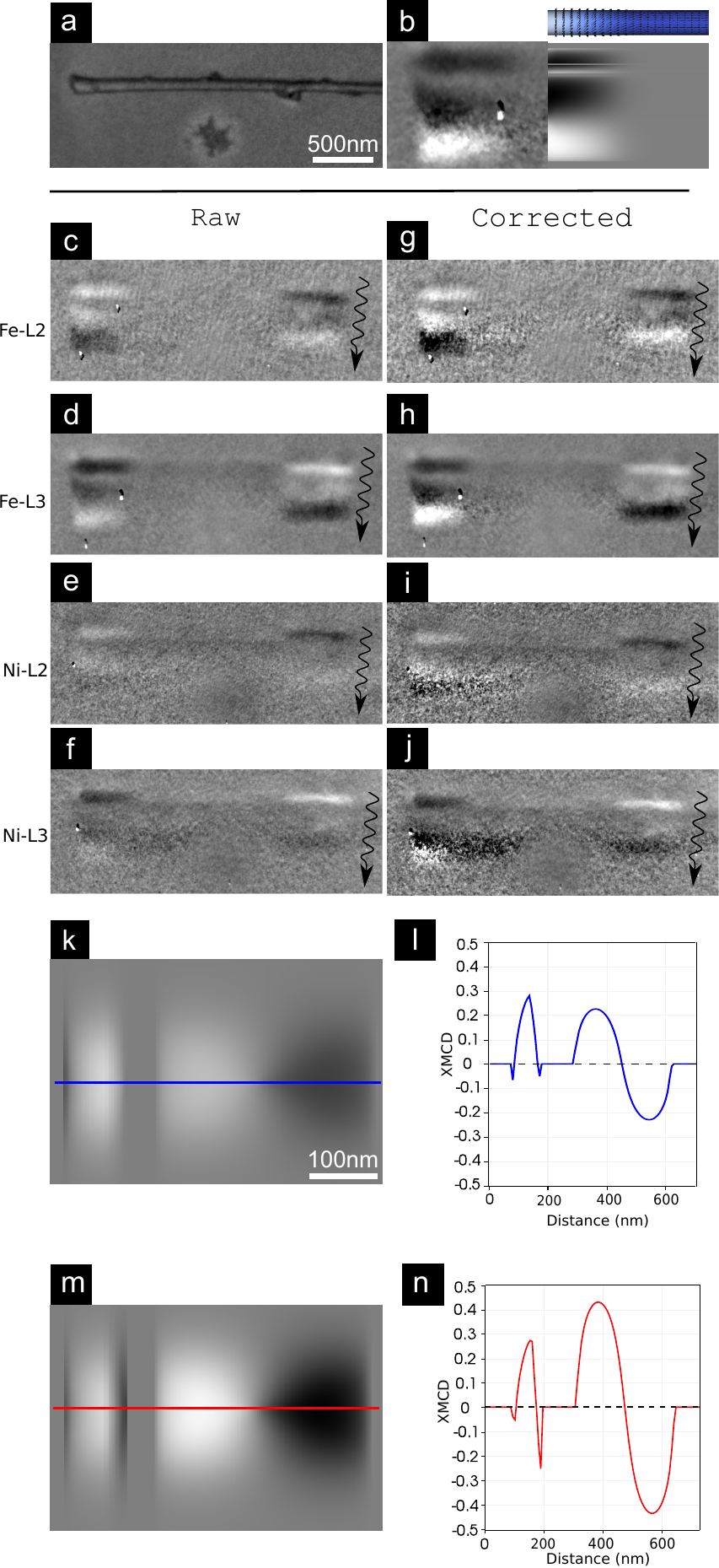}%
    	\caption{\textbf{Role of the absorption coefficient on the contrast.} (a) LEEM image of a wire with diameter $\lengthnm{120}$ (b) left end of the wire, with both experimental and simulated XMCD contrast at the Fe L3 edge, the latter revealing a curling micromagnetic configuration. (c-f) experimental XMCD contrast of this wire. (g-j) are the XMCD contrast computed from the same experimental XAS, however from which the background level has been removed.  Simulated XPEEM images (k,m) and their profiles (l,n) for a \unit[90]{\nano\meter}-diameter wire with a Bloch point wall and different absorbtion coefficients~(see \tabref{tab-coef-fig}). For (c,d,g,h) the contrast is \unit[6]{\%}, it is \unit[9]{\%} for (f,j) and \unit[5]{\%} for (e,i).}
  	\label{fig-inversion-expe}
  \end{center}
\end{figure}

Besides domain walls, three-dimensional non-uniform distributions of magnetization are also expected at diameter modulations and at the ends of wires, driven by the reduction of magnetostatic energy. Curling of magnetization around the wire axis has been predicted at such locations \cite{bib-LAN2009, bib-ALL2009}, however remaining elusive experimentally so far. What has been reported are hints for the spread of charges, which however could also be argued to take the form of other flux-closing structures\cite{bib-WAN2008a,bib-VOC2014}. Shadow PEEM again provides a direct proof for the existence of such buried structures. \subfigref{fig-inversion-expe}{(c-f)} shows the XMCD contrast of a wire at various absorption edges. A close-up view of the end of the wire is displayed in \subfigref{fig-inversion-expe}{b}, along with a simulation derived from a curling end domain. Thanks to the comparison we can formally identify the contrast at the end as arising from an orthoradial curling structure.

\afterpage{\clearpage}

\begin{table}
  \begin{center}
      \caption{The absorption coefficient values used for the simulations in \protect{\figref{fig-inversion-expe}}, (k-n). }
      \label{tab-coef-fig}
      \begin{tabular}{ccc}
    \hline\hline
    			&(k,l) &   (m,n)  \\\hline
    $\mu_{-}$ ($\reciprocal{\nano\meter}$) & 0.018 & 0.036 \\
    $\mu_{+}$ ($\reciprocal{\nano\meter}$) & 0.01 & 0.02 \\\hline\hline
  \end{tabular}
  \end{center}
	\end{table}
\section{Discussion on contrast}\label{sec-discussion}
\label{sec-contrast} In this section we discuss in more detail several instrumental aspects specific to the shadow imaging geometry, which have an impact on the magnetic contrast or spatial resolution. Of special importance for the shadow imaging of 3D objects are the plane of focus, the start voltage (STV, a voltage bias applied to the sample and which determines the electron kinetic energy), the microscope background level, the photon energy, and the Fresnel diffraction of X-rays.

\subsection{Microscope settings}
\label{microscope} 
\begin{figure}
	\begin{center}%
	\includegraphics{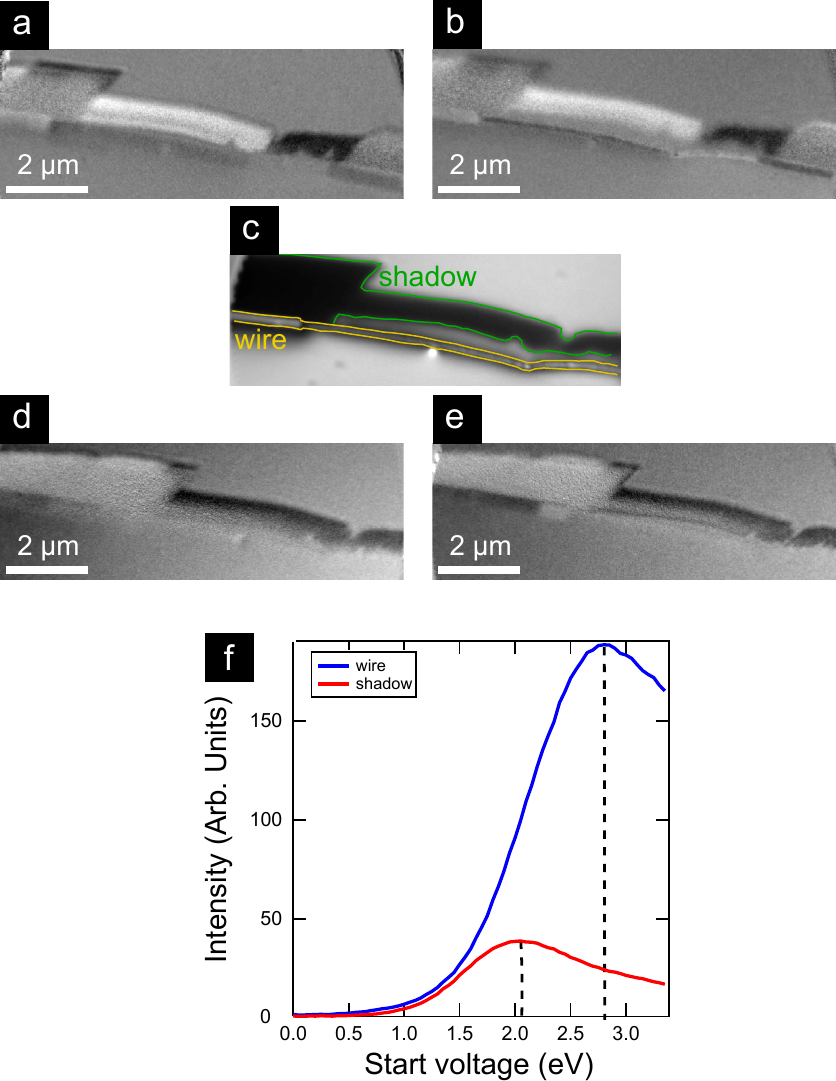}%
 \caption{\textbf{Role of the plane of focus and the start voltage.} XMCD contrast taken at the Fe L3 edge and with start voltage $\unit[2.1]{eV}$, with focus (a)~on the surface and (b)~on the wire. (c) is the XAS, superimposed with wire and shadow areas as a guide to the eye. (d)~XMCD with a focus on the surface, and start voltages $\unit[2.0]{eV}$ and (e)~$\unit[2.8]{eV}$.  (f)~XAS on the wire and the shadow versus the start voltage.}%
\label{fig-focus-stv}%
 	\end{center}
\end{figure}

While a rather flat surface may be entirely set close in focus, the case of three-dimensional objects lying on a surface is different. The depth of focus of the instrument is several micrometers, large enough so that the top of the wire and the supporting surface may both be in focus. In practice however this could not be achieved, which we understand as resulting from the wire curvature acting as a lens. For each image, one may thus decide to set the focus anywhere between the top surface of the wire, and the supporting surface. For instance, setting the focus on the supporting surface has a dramatic effect on loosing sharpness and therefore XAS and XMCD contrast on the wire, due to its small lateral size\bracketsubfigref{fig-focus-stv}{a,b}. Blurring effects are decreased upon increasing the start voltage, for reasons described below.

Second, electrons are extracted into the imaging column with a voltage $\unit[18]{\kilo eV} -$ STV. The start voltage (STV) is an additional bias, which is related to the electron kinetic energy (with an additional offset due to work function difference between sample and the $\mathrm{LaB}_6$ source used as reference for the energy scale). The non-planar wires we use create a non-uniform potential profile of the surface, which distorts the emitted electron wave. The lower the electron energy the more pronounced this distortion is. That may explain why the image quality both in the wire and in the shadow is better in \subfigref{fig-focus-stv}{e} than \subfigref{fig-focus-stv}{d}.

A fine tuning of the start voltage may also be used to enhance the signal originating from either the wire surface or the shadow. Indeed the materials giving rise to photoemission are different (here permalloy for the magnetic object, and Si for the supporting surface), as well as their capping, so the energy distribution and yield of secondary photoelectrons are different\bracketsubfigref{fig-focus-stv}{f}. In the present case a lower start voltage~($\approx\unit[2.0]{eV}$) maximized the number of electrons emitted from the shadow, whereas a higher start voltage ($\approx\unit[2.8]{eV}$) resulted in a higher intensity emitted from the wire surface\bracketsubfigref{fig-focus-stv}{e}. XAS being the measure of emitted electrons, these effects of start voltage are directly transferred to the XAS image. As XMCD is a difference normalized to a sum, its magnitude should not depend on the number of electrons and thus be insensitive to the choice of start voltage.  However, as will be argued in \secref{sec-bcglevel}, in practice a lower number of emitted electrons reduces the XMCD signal with respect to the computed value, so the start voltage also has an impact on the relative level of contrast on the wire versus the shadow.

\subsection{Photon energy}\label{sub-sec-photon}
For 3d ferromagnetic metals the photon energy needs to be tuned close to the maximum of an L3 edge to maximize dichroism. Attention should be paid to the fact that event a slight surface oxidation, induces a sub-structure in the L3 peak. This slightly shifts the XAS maximum with respect to a metallic spectrum\bracketsubfigref{fig-xas-xmcd}c, while XMCD remains maximum at the location of the metal peak. Accordingly, we found more precise to use an absorption spectrum taken in the shadow to set the working photon energy, as this probes the bulk of the wire, with no oxidation. Notice that this choice also maximizes the dichroic contrast at the surface of slightly-oxidized wires. In practice, we worked at the Fe L3 edge, which despite the low concentration of Fe, proved to yield a larger contrast than the Ni L3 edge, both at the surface and in the shadow of the wires. This will be addressed in \secref{sec-bcglevel}.

\afterpage{\clearpage}
We showed in \secref{sec-test-cases} that a positive XMCD contrast on the wire should be associated with a negative XMCD contrast in the shadow: a larger absorption and thus loss of photons of a given helicity is associated with an enhanced number of emitted electrons. The number of photons going through the magnetic object depends on the dimensionless quantity $d\mu_{\pm}$, with $d$ its the depth, here the diameter of the wire. The effect of varying this quantity is illustrated by a movie of the wire and its shadow upon ramping the photon energy from below Fe L edges to above Ni L edges (see supplementary materials). For large enough $d\mu_{\pm}$, the imbalance of photons with opposite helicities may be sufficient on the backside of the wire to outweigh dichroism, so that the contrast on the wire may  be reversed from front to back side even for the same direction of magnetization. This expectation is clear in \figref{fig-test-cases}, and had been reported experimentally previously\cite{bib-STR2012b}. To illustrate the expected impact of $d\mu_{\pm}$ on the front, back and shadow contrast, \subfigref{fig-inversion-expe}{(k,m)} illustrates the XMCD contrast of a Bloch-point wall in the same wire with two different sets of absorption coefficients (\tabref{tab-coef-fig}). \subfigref{fig-inversion-expe}{(l,n)} presents their profiles. For higher $d\mu_{\pm}$, there is a contrast inversion at the back side of the wire~[\subfigref{fig-inversion-expe}{(k,m)}]. In practice, combining images of the same area varying $\mu$ may be useful to refine the analysis. To illustrate this, \subfigref{fig-inversion-expe}{(c-f)} show XMCD images of the same wire and measured at the $\mathrm{L}2$ and $\mathrm{L}3$ edges of both Fe and Ni. A large value of $\mu$ and of the difference $\mu_+ - \mu_-$ is potentially an advantage for areas where the through-thickness is moderate, to maximize contrast, while a low value of $\mu$ is potentially a better choice in the case of long distances traveled through the magnetic object, to keep a reasonable signal-over-noise ratio~(see discussion in \secref{sec-resolution-noise}).

\subsection{Background level in PEEM imaging}\label{sec-bcglevel}
On a theoretical basis the contrast in the shadow could reach arbitrarily high values for high $\mu d$, which however comes at the expense of much reduced intensity. This is the principle of some polarizers, for example for the helicity of X-rays \cite{bib-KOR1997} or spin of electrons \cite{bib-PAP1991}. Aside from obvious issues arising from the signal-over-noise ratio, we found out that in practice an instrumental effect limits the contrast. \figref{fig-interferences} shows the level of XAS on a broad wire and its shadow. For this broad wire $\mu_{\pm} d\gg1$, so that the intensity in the shadow should be vanishingly small. To the contrary, although the intensity reaches a plateau inside the shadow, it remains close to $\unit[7]{\%}$ of the intensity over the free supporting surface\bracketsubfigref{fig-interferences}d. This intensity is not related to the background electronic level of the camera, which is already subtracted from the images. Instead, it reflects electrons that truly impinge on the detector. The physical origin of this background is not straightforward, as it was found to be only weakly affected by changing settings of the LEEM. In particular the field-of-view aperture was found to be unrelated, although rejecting electrons from the imaging column arising from outside the field of view, to avoid their incoherent contribution to the image. The contrast aperture, affecting the angular collection of the microscope, did not have a sizeable impact either. Thus although its origin is not clear, an (a priori) helicity-independent background intensity reduces the computed XMCD as its difference in the numerator is zero, while its sum in denominator is non-zero. If the background intensity $I_{\mathrm{e},\mathrm{b}}$ is known, then a more accurate view of the true XMCD is achieved by computing: $I_{\mathrm{XMCD,0}} = (I_{\mathrm{e},\sigma_{-}} - I_{\mathrm{e},\sigma_{+}}) / (I_{\mathrm{e},\sigma_{-}} + I_{\mathrm{e},\sigma_{+}} - 2I_{\mathrm{e},\mathrm{b}})$. This has been done in \subfigref{fig-inversion-expe}{(g-j)}. It is striking that the contrast in the shadow is enhanced, as expected from theory. These comparisons also illustrate that working at the Fe edges yields in practice a higher contrast than at the Ni edges, whereas a similar contrast would be expected for permalloy as computed from the tabulated absorption coefficients (Table \ref{tab-coeffabs}). This is explained simply by the existence of the background level.

\subsection{Scattering effects}

In general, interaction of X-rays with matter can be described via the complex atomic scattering factors. The real and imaginary parts give rise to the Faraday rotation of the photon polarization vector and to magnetic dichroism, respectively, as the photon beam propagates through the magnetic material. The two are related by the Kramers-Kronig transformation, and they are comparable in magnitude at the Fe L3 edge \cite{bib-KOR2000}.

\begin{figure}
	\begin{center}%
		\includegraphics{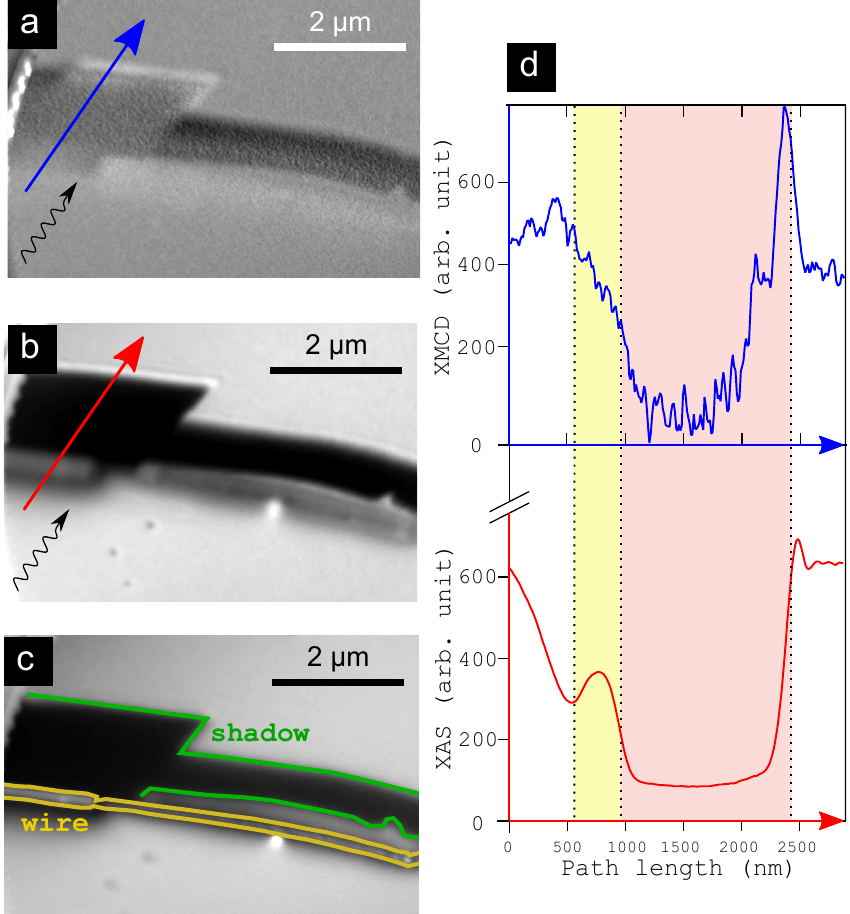}%
    \caption{\textbf{Scattering effects.} Wire image of the intensity for (a) the XMCD and (b) the XAS. (c) shows the wire and shadow zones on the XAS images. The plot (d) presents the intensity at the cross-sections shown in (a) and (b). Dotted lines are guides to the eyes and limit wire~(left) and shadow areas~(right).}%
    \label{fig-interferences}%
\end{center}
\end{figure}

Until now, we have considered only the X-ray absorption coefficient, which is proportional to the imaginary part of the forward scattering amplitude via the optical theorem \cite{bib-JAC1975}. Instead, as we noted above, a proper treatment should include the full scattering process. Indeed, intensity oscillations near the shadow edge are visible in \figref{fig-interferences} due to Fresnel diffraction from the wire. Furthermore, the Fresnel fringes also show a dichroic signal. The sign of this dichroic signal is opposite to that observed within the shadow, as expected from the inverted absorption signal in transmission.

Nevertheless, the shadow (or the substrate) is in the very near field of the wire, and coherent scattering effects are limited to the shadow edge. The Fraunhofer region at this wavelength (about \unit[1.8]{\nano\meter}) and for a wire diameter below \unit[100]{\nano\meter} does not before a few tens of microns. Therefore, our analysis relating the dichroism within the shadow to the absorption coefficient is valid except at the very edge of the shadow\cite{bib-LOC2010b}.

\section{Discussion on spatial resolution}
\label{sec-resolution}

In this last section we discuss spatial resolution effects specific to shadow imaging.

\subsection{Electric-field related distortion}
First, it is known that LEEM images of non-planar surface are distorted\cite{bib-SCH2002}. The physical phenomenon is that secondary electrons escape the material perpendicular to the local surface, on the average. Thus their trajectory is curved through the extraction electric field, the curvature of the object providing the same effect as a lens\bracketsubfigref{fig-scheme-coll}{b}. Besides curvature, the complex topography of the sample acts as a cathode for the accelerating voltage. Therefore the accelerating voltage is not uniform across the surface. This creates a significant distortion to the outgoing low-energy electron wave. The LEEM image of convex and topographically-complex objects such as the wires considered here, is therefore expected to display a larger size than the real object. Obviously this phenomenon is largely absent in the shadow, as the trajectory of photons is only weakly affected by the circular shape of the wires, due to the optical index being very close to unity. For wires lying perfectly on the supporting surface this effect cannot be checked, because the direct and shadow contrasts overlap. In a few cases a gap was found between a wire and the surface, large enough to separate the direct and shadow areas\bracketsubfigref{fig-bpw}{b}. For such cases the apparent width of the wire deduced from XAS images was indeed about $\unit[50]{\%}$ larger on the wire than in the shadow. To minimize systematic errors, the figures for wire diameter mentioned in the manuscript and used for simulations, have always been those deduced from the shadow and deconvoluted from the expected $\lengthnm{30}$ experimental resolution.

\subsection{Signal-over-noise ratio in the shadow}
\label{sec-resolution-noise}

\begin{figure}
	\begin{center}%
		\includegraphics{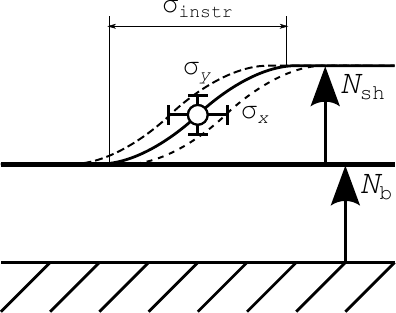}%
    \caption{\textbf{Signal-over-noise ratio}. Schematic to describe the notations used in the calculation (see text).}%
    \label{fig-incertitudes}%
\end{center}
\end{figure}

Not only is the shadow more faithful as just discussed, but it may promise for an increase of spatial resolution by a factor $1/\sin(\angledeg{16})\approx3.6$ thanks to the projection with a rather grazing incidence. This would bring the spatial resolution along one direction below $\lengthnm{10}$. A practical limitation for this gain is the lower number of electrons collected in the shadow, degrading the signal-over-noise ratio as estimated below. Let $N$ be the number of electrons emitted from the supporting surface under direct illumination, per given time and area. $N_\mathrm{b}$ and $N_\mathrm{sh}$ are similarly the number of electrons contributing to the background level, and those contributing to the shadow and related to photons transmitted through the wire, again per unit area and time. In our case $N_\mathrm{b}/N\approx0.07$ and $N_\mathrm{sh}/N\approx\exp(-\mu d)$, with $d$ the diameter of the wire. The shortest spatial variation that can be expected on the detector is the instrumental resolution $\sigma_\mathrm{instr}$, resulting in a slope $f'=t N_\mathrm{sh}/\sigma_\mathrm{instr}$ with $t$ the averaging time. When analyzing experimental data, the possible error on lateral resolution $\sigma_x$ resulting from the vertical error bar $\sigma_y$ is such that $\sigma_y/\sigma_x=f'$~(see \figref{fig-incertitudes}). We thus have: $\sigma_x=\sigma_\mathrm{instr}(\sigma_y/tN_\mathrm{sh})$. Taking into account that $\sigma_y=\sqrt{tN_\mathrm{b}}+\sqrt{tN_\mathrm{sh}}$, one finally gets:

 \begin{equation}\label{eqn-statistics-in-shadow}
   \sigma_x = \frac{\sigma_\mathrm{instr}}{\sqrt{tN_\mathrm{sh}}}
       \left({1+\sqrt{\frac{N_\mathrm{b}}{N_\mathrm{sh}}}}\right)
 \end{equation}
In the absence of background level, \eqnref{eqn-statistics-in-shadow} boils down to the usual statistics: $\sigma_x=\sigma_\mathrm{instr}/\sqrt{tN_\mathrm{sh}}$. Thus in theory an image of quality similar to that outside the shadow with integration time~$t_0$, could be obtained at the expense of an increase in integration time up to $t_\mathrm{sh}$ such that $\sqrt{t_\mathrm{sh}N_\mathrm{sh}}=\sqrt{t_0 N}$, so with an increase of time $N/N_\mathrm{sh}=\exp(\mu d)$. This ratio is of the order of $10^2 -10^4$ for a wire with a diameter of \unit[100]{\nano\meter} at the Fe-L$3$ edge. However in the case of non-zero background level, fluctuations in $N_\mathrm{b}$ also contribute to the increase of $\sigma_y$ and thus of $\sigma_x$. If $N_\mathrm{sh}$ becomes small compared to $N_\mathrm{b}$, based on \eqnref{eqn-statistics-in-shadow} the time required to get an image of similar quality is $t_\mathrm{sh}=t_0 (N/N_\mathrm{sh})^2(N_\mathrm{b}/N)$, thus with now the power law $\exp(2\mu d)$. For a diameter \unit[100]{\nano\meter}, the power law is now proportional to $10^4-10^8$ at the Fe L$3$ edge which becomes prohibitively large. It is the same effect of limited statistics, which limits the signal-over-noise ratio of dichroism in the shadow. While in \secref{sub-sec-photon} we saw that dichroism should asymptotically reach 100\% in deep shadows~($\mu d\gg 1$), one can show that in practice the signal-over-noise is maximum for $\mu d\approx 1$.

\section{Discussion and Conclusion}
\label{sec-conclusion}

At this stage a comparison may be made with the transmission X-ray microscope (TXM). Indeed both TXM and shadow-PEEM allow for probing the volume magnetization integrated along the photon beam. An advantage of TXM is its all-photon basis, making it easily compatible with applied magnetic fields. Also, the sample may be rotated to some extent, gaining information on different directions of magnetization or integration. On the reverse, shadow-PEEM provides the combination of surface and volume information, which may be crucial to solve complex three dimensional magnetization distributions. The potential increase of spatial resolution is also unique. Experiments may even be designed with magnetic objects tilted on purpose to a chosen angle to make the best use of this gain.

To conclude, we have discussed quantitatively physical and instrumental features specific to shadow-PEEM imaging of three-dimensional objects lying on a supporting surface. We have considered in more detail XMCD imaging and simulation of the expected contrast from micromagnetic simulations. This technique uniquely provides the combination of surface and volume sensitivity in the signal measurement, with an enhanced XMCD contrast and several-fold gain in spatial resolution along the beam direction for the latter. Several effects mentioned need however to be considered to extract true spatial and contrast information such as plane of focus, extraction voltage, electric field distortion and electron background level. While we illustrated the method with experiments we performed on cylindrical nanowires, it can be applied to any object of size $\unit[10\mathrm{-}200]{\nano\meter}$ imaged by shadow XMCD-PEEM, such as those already reported\cite{bib-KIM2011b,bib-STR2012b,bib-SAN2012,bib-ZAB2013}, provided that material parameters such as exchange stiffness and magnetization are known, to perform reliable micromagnetic simulations. In particular, sufficient geometrical information about the sample remains required, as its projected shadow does not characterize fully its shape.

\section*{Acknowledgements}

We thank A.~Sala (ELETTRA) for assistance in the XMCD-PEEM measurements,  J.~Vogel, J. -M Tonnerre and S.~Pizzini~(Institut NEEL), and R.~Belkhou~(Synchrotron Soleil) for useful discussions. The research leading to these results has received funding from the European Community's Seventh Framework Programme (FP7/2007-2013) under grant agreements n$\deg$312284 and 309589.

\bibliographystyle{apsrev}
\bibliography{Fruche6}

\end{document}